\documentclass[12pt]{iopart}
\newcommand{\p}{\partial}

\newcommand{\ppkz}{\frac{\partial}{\partial k_z}}

\usepackage{graphicx}

\begin{document}

\title[Strange and charm quark-pair production in strong non-Abelian field]{Strange and charm quark-pair production\\ in strong non-Abelian field}
\author{P\'eter L\'evai}
\address{KFKI RMKI Research Institute for Particle and Nuclear Physics, \\
P.O. Box 49, Budapest 1525, Hungary}
\ead{plevai@rmki.kfki.hu}

\author{Vladimir V. Skokov}
\address{Bogoliubov Laboratory of Theoretical Physics, 
Joint Institute for Nuclear Research, \\
Dubna, 141980, Russia }
\address{GSI, Planckstra\ss{}e 1, D-64291 Darmstadt, Germany }
\ead{V.Skokov@gsi.de}


\begin{abstract}

We have investigated strange and charm quark-pair production in 
the early stage of heavy ion collisions. Our kinetic model is based on
a Wigner function method for fermion-pair production in strong non-Abelian
fields. 
To describe the overlap of two colliding heavy ions  we have 
applied the time-dependent color field with a pulse-like shape. 
The calculations have been performed
in an SU(2)-color  model with finite current quark masses.
For strange quark-pair production the obtained results are close to the 
Schwinger limit, as we expected. 
For charm quark the large inverse temporal width of the field  pulse, 
instead of the large charm quark mass,  determines the efficiency of the
quark-pair production. 
Thus we do not 
observe
the expected suppression of charm quark-pair production connecting to
the usual Schwinger-formalism, but our calculation results in a relatively
large charm quark yield. 
This effect appears in Abelian models  as well, 
demonstrating that  particle-pair production for fast varying non-Abelian 
gluon field strongly deviates from the Schwinger limit for charm quark. 
We display our results on
number densities for light, strange, charm quark-pairs, and different
suppression factors  as the function of 
characteristic time of acting chromo-electric field.
\end{abstract}



\section{Introduction}

In the transport models 
theoretical descriptions of particle production in high energy $pp$
collisions are based on the introduction of chromoelectric flux tube
('string') models, where these tubes are connecting quark and
diquark constituents of colliding protons~\cite{FRITIOF}.
However, at RHIC and LHC energies the string density is expected to be
so large that a strong collective gluon field will be formed in the whole
available transverse area. 
Furthermore, the gluon number will be so high that a classical gluon field
as the expectation value of the quantum field can be considered
in the reaction volume~\cite{Gyul97,Topor05}.
We have investigated quark-pair production and determined particle spectra  in  
time-dependent external U(1) and SU(2) chromo-electric fields~\cite{SkokLev05,SkokLev07}.
In this paper, we describe strange and charm quark-pair production, and
make a calculations of corresponding  suppression factors for SU(2) gauge
field. 
The results of solving similar problem for U(1) gauge field 
can be found in~\cite{Prozorkevich:2004yp}.

An alternative approach, that takes into account space inhomogeneities,  was 
considered in~\cite{Lappi,GelisLappi}. However, it is worthwhile  
to mention that in contrast to the main idea of ~\cite{Lappi,GelisLappi}, 
where pairs production is directly calculated by numerical integration 
of a Dirac equation,  our approach based on solving a kinetic equation
for an ``observable'' Wigner function (or, finally,  distribution function)
providing to a considerable extent  an intuitive insight to the physical problem.  
The  next  advantage of the current approach that it is not so highly computer
demanded, thus allows to obtain detailed information about created particles.

\section{The kinetic equation for the Wigner function}

The equation of motion for color Wigner function in gradient 
approximation reads \cite{HeinzOchs,Prozor03}:
\begin{eqnarray}
&&\partial_t W+ \frac{g}{8}\frac{\partial}{\partial k_i}
\left( 4\{W,F_{0i}\} 
 + 
2\left\{F_{i\nu},[W,\gamma^0 \gamma^\nu]\right\}-
\left[F_{i\nu},\{W,\gamma^0 \gamma^\nu\}\right] \right)=\nonumber\\
&&=ik_i \{\gamma^0 \gamma^i,W\}-im[\gamma^0,W] +ig \left[A_i\,, 
[\gamma^0 \gamma^i ,W]\right] . \ \ \ 
\label{W}
\end{eqnarray}

The color decomposition of the Wigner function with SU($N_c$) generators in fundamental representation 
is given by 
\begin{equation}
W = W^s + W^a t^a, \,\, \ \ \ a = 1,2,..., N_c^2-1 \ ,
\label{color_decomposition}
\end{equation}
where $W^s$ is the color singlet part and $W^a$ is the color multiplet components. 
It is also convenient to perform spinor decomposition separating scalar $a$, vector
$b_\mu$, tensor $c_{\mu\nu}$, axial vector $d_\mu$ and pseudo-scalar parts $e$:
\begin{equation}
W^{s|a} = a^{s|a} + b^{s|a}_\mu \gamma^\mu + c^{s|a}_{\mu\nu} \sigma^{\mu\nu} +
d^{s|a}_\mu \gamma^\mu \gamma^5 + i e^{s|a} \gamma^5.
\label{Clifford_decomposition}
\end{equation}
The asymmetric tensor  components  of the Wigner function can be  
decompose into axial and polar vectors $c_1^j=c^{j0}$ and 
$c_2^j=\frac{1}{2}\epsilon^{0 \omega\rho j}c_{\omega\rho}$ correspondingly. 

 \section{Kinetic equation  with SU(2) color isotropic external field}
After decomposition the equations for the Wigner function in the case of 
pure longitudinal external SU(2) color field, $\bi{A}^a=(0,0,0,A_z^a(t))$, with fixed color direction 
$A^a_z = A^\diamond_z n^a$, where  $n^a n^a=3$ and $\p_t n^a=0$~\cite{SkokLev07}, 
 we obtain the following  system of equations for  singlet components
\begin{eqnarray}
&&\p_t a^s + \frac{3 g}{4} E^\diamond_z  \ppkz  a^\diamond  = - 4 \bi{k} \bi{c}_1^s,    \\
&&\p_t e^s + \frac{3 g}{4} E^\diamond_z  \ppkz    e^\diamond  = - 4 \bi{k} \bi{c}_2^s - 2  m d^s_0,\\
&&\p_t b^s_0 + \frac{3 g}{4} E^\diamond_z   \ppkz  b^\diamond_0 =  0,\\
&&\p_t \bi{b}^{s} + \frac{3g}{4} E^\diamond_z \ppkz \bi{b}^\diamond =  2   [ \bi{k} \times \bi{d}^s ]  + 4  m \bi{c}_1^s,\\
&&\p_t d^s_0 + \frac{3g}{4} E^\diamond_z \ppkz   d^\diamond_0  =  2  m  e^s, \\
&&\p_t \bi{d}^s + \frac{3g}{4} E^\diamond_z \ppkz   \bi{d}^\diamond =  2 [\bi{k} \times  \bi{b^s}], \\
&&\p_t \bi{c}_1^s  + \frac{3g}{4} E^\diamond_z \ppkz \bi{c}_1^\diamond  = a^s \bi{k} - m \bi{b}^s,\\
&&\p_t \bi{c}_2^s  + \frac{3g}{4} E^\diamond_z \ppkz \bi{c}_2^\diamond  =  e^s \bi{k};
\end{eqnarray}
and multiplet components 
\begin{eqnarray}
&&\p_t a^\diamond + g E^\diamond_z \ppkz  a^s   = - 4 \bi{k} \bi{c}_1^\diamond, \\
&&\p_t e^\diamond + g E^\diamond_z \ppkz e^s  = - 4 \bi{k} \bi{c}_2^\diamond- 2m d^\diamond_0, \\
&&\p_t b^\diamond_0 + g E^\diamond_z \ppkz  b^s_0  = 0,\\
&&\p_t \bi{b}^\diamond + g E^\diamond_z \ppkz \bi{b}^s 
=  2 [ \bi{k} \times  \bi{d}^\diamond ]  + 4 m \bi{c}_1^\diamond,\\
&&\p_t d^\diamond_0 + g E^\diamond_z \ppkz  d^s_0 \delta^{b c}  
= 2 m e^\diamond,\\
&&\p_t \bi{d}^\diamond + g E^\diamond_z \ppkz \bi{d}^s 
= 2 [ \bi{k} \times \bi{b}^\diamond ],\\
&&\p_t \bi{c}_1^\diamond + g E^\diamond_z \ppkz \bi{c}_1^s
= a^\diamond \bi{k}  - m \bi{b}^\diamond,\\
&&\p_t \bi{c}_2^\diamond + g E^\diamond_z \ppkz \bi{c}_2^s 
= e^c \bi{k}, 
\end{eqnarray}
where SU(2) triplet components of the Wigner function are defined 
by $a^a  = a^\diamond n^a$.

The distribution function of quarks (antiquarks)
is defined by the components 
$a^s, \bi{b}^s$~\cite{SkokLev07}:
\begin{equation}
f_q(\bi{k},t) = f_{\bar{q}}(-\bi{k},t)  = \frac{m a^s(\bi{k},t) +  \bi{k} \, \bi{b}^s(\bi{k},t)}
{\omega(\bi{k})}  + \frac{1}{2}, \ \   {\omega(\bi{k})} = \sqrt{\bi{k}^2+m^2}.
\label{DF}
\end{equation}
Thus to obtain the  
 quark distribution 
function, the  scalar $a$,  vector $b_\mu$, axial vector $d_\mu$, and axial tensor $c_1^{\mu}$  components 
of the Wigner  function are required, only. 

The initial conditions for the Wigner function in vacuum reads \cite{SkokLev07}:
\begin{equation}
a^s = -\frac{1}{2} \frac{m}{\omega}, \quad \bi{b}^s =  -\frac{1}{2} \frac{\bi{k}}{\omega}, 
\label{Initial}
\end{equation}
and zero initial conditions for the rest components of Wigner function.
Considering symmetry of  
initial condition  and performing the  vector decomposition, 
\begin{equation}
\bi{v} = v_z \bi{n} + v_\perp \frac{\bi{k}_\perp}{k_\perp} 
+ v_x [ \bi{n} \times \frac{\bi{k}_\perp}{k_\perp} ] \ , 
\end{equation}
we obtain the following equations for
singlet components ($\bi{c}=\bi{c}_1$ 
to simplify reading):
\begin{eqnarray}
\label{final_b}
&&\p_t a^s + \frac{3 g}{4} E^\diamond_z  \ppkz  a^\diamond  = - 4 (k_z c_z^s + k_\perp c_\perp^s),    \\
&&\p_t b_z^s + \frac{3g}{4} E^\diamond_z \ppkz b_z^\diamond = 2 k_\perp  d_x^s   +  4  m c_z^s,\\
&&\p_t b_\perp^s + \frac{3g}{4} E^\diamond_z \ppkz b_\perp^\diamond = - 2 k_z  d_x^s   +  4  m c_\perp^s,\\
&&\p_t d_x^s + \frac{3g}{4} E^\diamond_z \ppkz d_x^\diamond = 2( k_z  b_\perp^s - k_\perp b_z^s ),\\
&&\p_t c_z^s  + \frac{3g}{4} E^\diamond_z \ppkz c_z^\diamond  = a^s k_z - m b_z^s,\\
&&\p_t c_\perp^s  + \frac{3g}{4} E^\diamond_z \ppkz c_\perp^\diamond  = a^s k_\perp - m b_\perp^s \ ;
\label{final_e}
\end{eqnarray}
and for multiplet components: 
\begin{eqnarray}
&&\p_t a^\diamond + g E^\diamond_z \ppkz  a^s   = - 4 (k_z c_z^\diamond + k_\perp c_\perp^\diamond), \\
&&\p_t b_z^\diamond + g  E^\diamond_z \ppkz b_z^s = 2 k_\perp  d_x^\diamond   +  4  m c_z^\diamond,\\
&&\p_t b_\perp^\diamond + g E^\diamond_z \ppkz b_\perp^s = - 2 k_z  d_x^\diamond   +  4  m c_\perp^\diamond,\\
&&\p_t d_x^\diamond + g  E^\diamond_z \ppkz d_x^s = 2( k_z  b_\perp^\diamond - k_\perp b_z^\diamond ),\\
&&\p_t c_z^\diamond  +  g E^\diamond_z \ppkz c_z^s  = a^\diamond k_z - m b_z^\diamond,\\
&&\p_t c_\perp^\diamond  + g E^\diamond_z \ppkz c_\perp^s  = a^\diamond k_\perp - m b_\perp^\diamond \ .
\end{eqnarray}
The  axial part of vector $b_x$ and tensor components $c_x$, longitudinal $d_z$ and transverse  $d_\perp$
parts of axial vector components do not contribute to the evolution of the distribution function and thus are not considered.

\section{Numerical results and discussions} 

In Ref.~\cite{SkokLev07} we have solved the above equations and described the time evolution
of the quark  distribution functions to obtain the longitudinal and transverse 
quark spectra. In contrast to Ref.~\cite{SkokLev07}, in the current paper we focus on the integrated particle yields. 

In the numerical calculation we have used the following parameters: 
the maximal string tension $E_0=1$ GeV/fm; 
coupling constant $g=2$;
the current quark masses $m_{u,d}=8$ MeV, $m_s=150$ MeV, $m_c=1200$ MeV 
for light, strange  and charm quark, respectively. 
The particle production is ignited by a pulse-like 
 field
$E_{z}^\diamond(t) = E_0 \cdot \left[ 1- \textrm{tanh}^2 (t/\tau) \right]$, 
which is characterized by the amplitude of the pulse $E_0$ and its temporal width $\tau$.  
In this treatment particles are produced and absorbed by the field pairwise. 

\begin{figure}[t]
\centerline{
\includegraphics[height=5.8truecm] {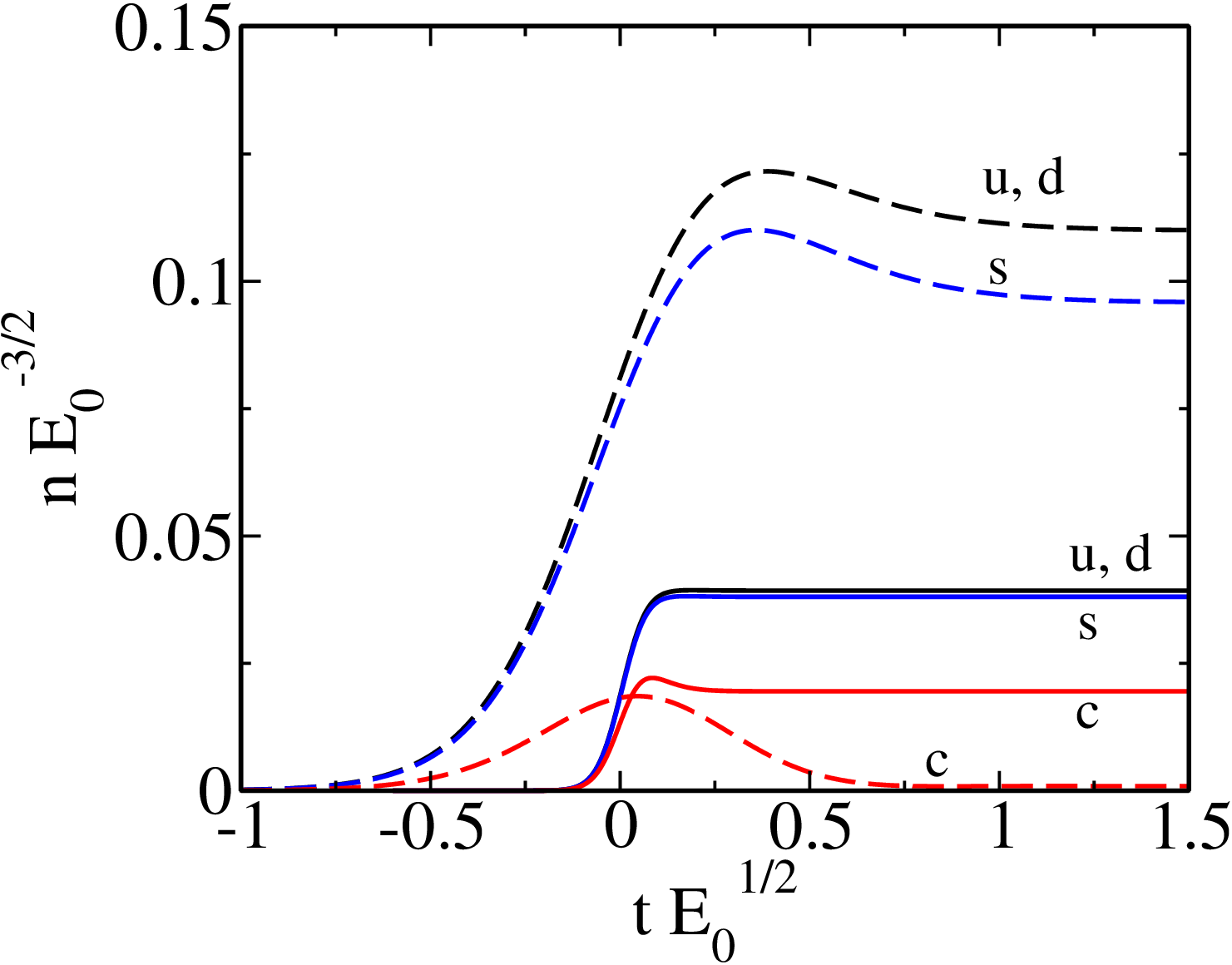}
\includegraphics[height=5.8truecm] {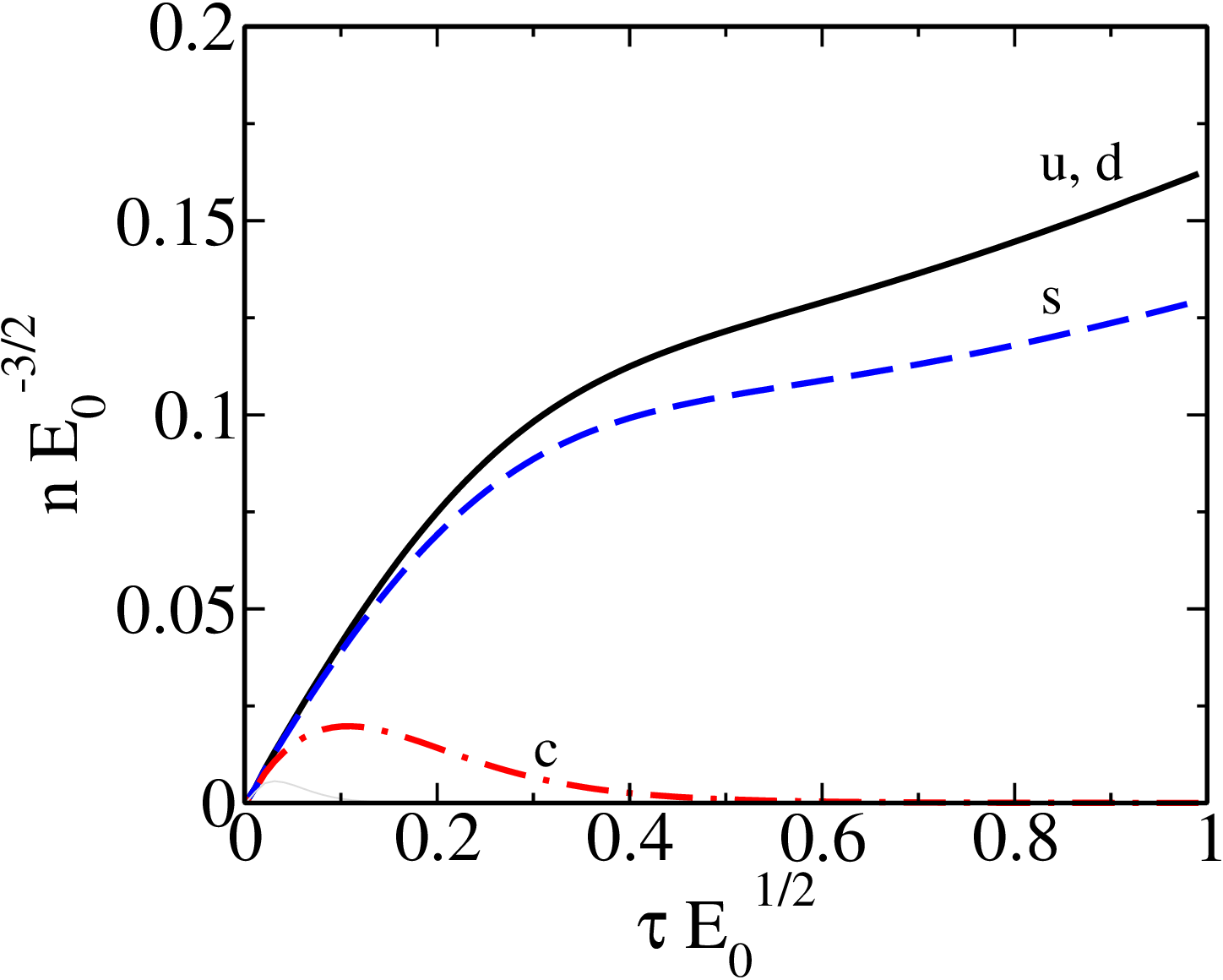}}
\caption{
Left panel: the total quark-pair number densities  for different flavours, $n_i(t)$,
as a function of time  $t$ 
for short pulse width $\tau E_0^{1/2}=0.1$ (solid lines) and long pulse width
$\tau E_0^{1/2}=0.5$ (dashed lines). Right panel:   
the total quark-pair number densities  at the  final state, $n(t \gg \tau)$,  
as a function of pulse width  $\tau$.
}
\label{nt}
\end{figure}

\begin{figure}[b]
\centerline{
\includegraphics[height=5.8truecm] {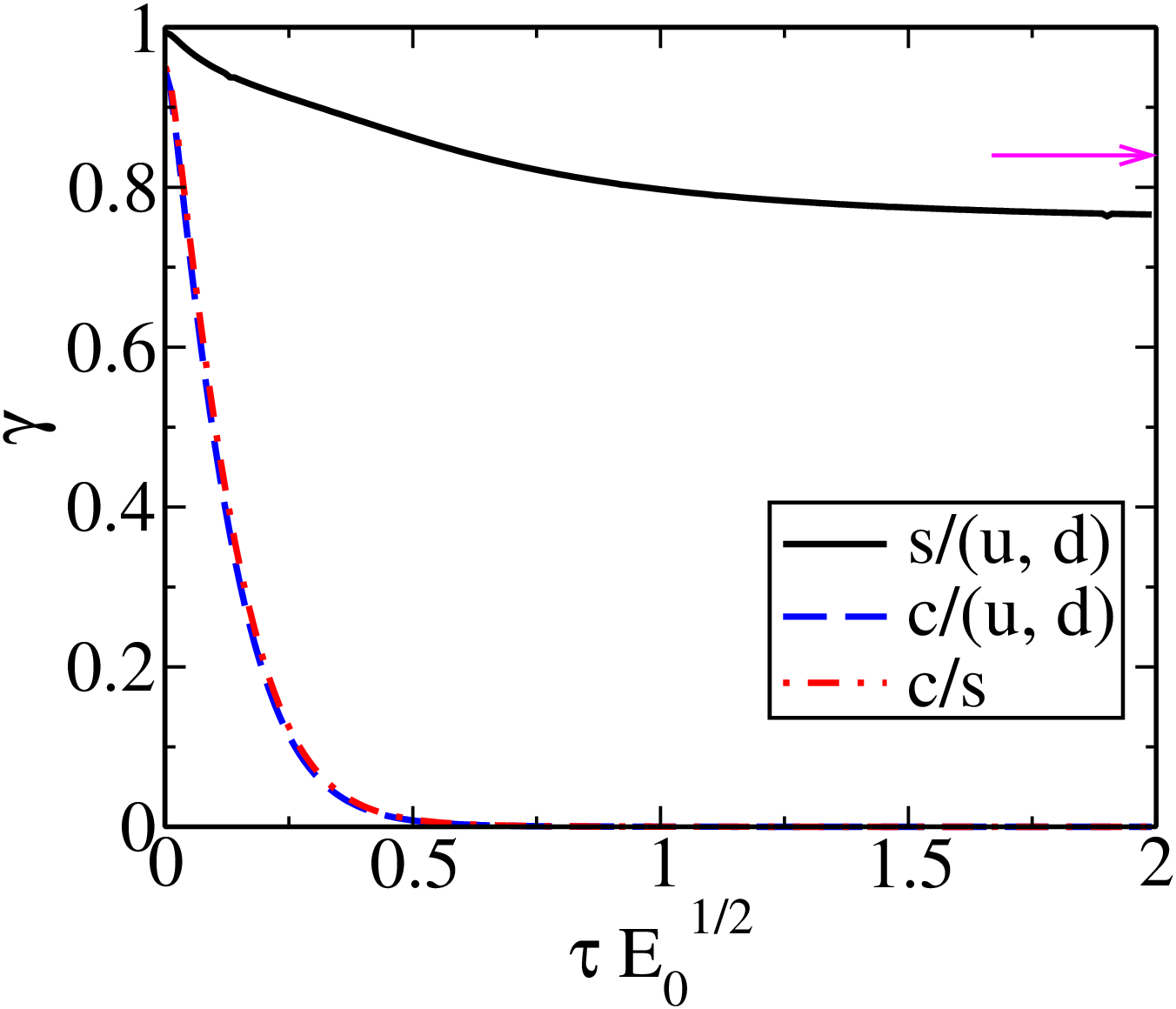} \hspace*{0.2truecm}
\includegraphics[height=5.8truecm] {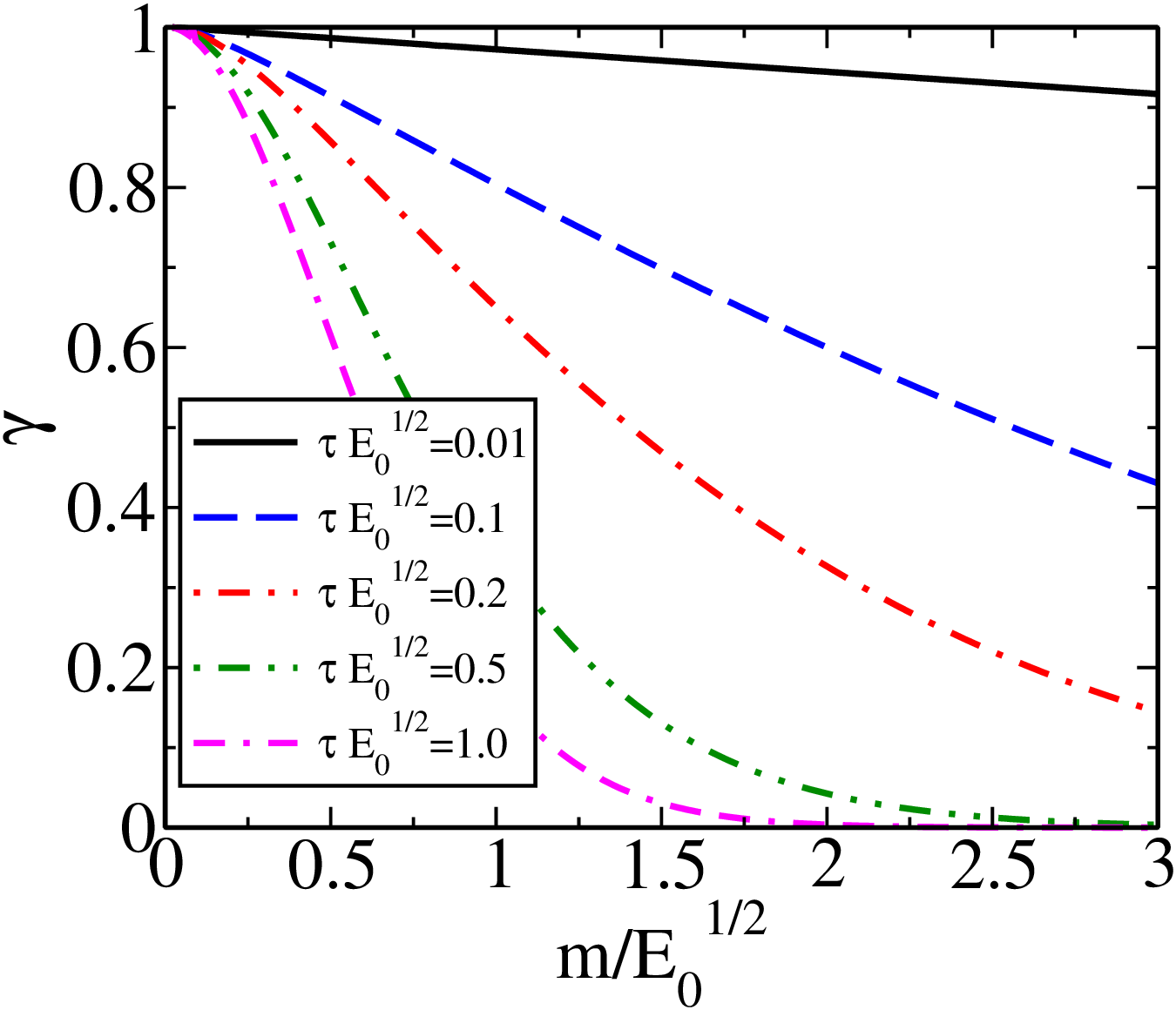}}
\caption{
Left panel: the pulse width, $\tau$, dependence of the suppression factor $\gamma$.
Arrow indicates the Schwinger limit for strangeness suppression factor.
Right panel: the quark mass, $m$, dependence of the suppression factor $\gamma$
at different pulse width. 
}
\label{gamma}
\end{figure}

The ratio of number densities of  heavy quark-pairs, e.g. strange,  to  light quark-pairs (u, d-quarks) is widely known as 
a suppression factor. In our model it is defined 
in the asymptotic future (c.f. ~\cite{schw51}), $t \gg \tau$, as 
\begin{equation}
\gamma = \lim_{t\to\infty} n_{\rm heavy}(t)/n_{\rm light}(t), 
\end{equation}
where $n_q(t)$ is the number density of corresponding quark-pairs given by 
\begin{equation}
n_q(t) =\nu \int  \frac{d^3k}{(2\pi)^3} f_q(\bi{k},t) 
\label{numberdens}
\end{equation}
with degeneracy factor $\nu = 2 ({\rm spin}) \times 2 ({\rm quark-antiquark}) \times N_c ({\rm color}) $.

In Fig. \ref{nt} the time evolution of quark-pair number densities, $n_i$, are displayed for different pulse widths,
$\tau E_0^{1/2}=0.1$ and 0.5. 
For short pulse width the quark-pair  number  densities are comparable with each other (solid lines). 
In this case the particle production happens during the whole evolution of the field.
In contrast to this, for long pulse, the number densities  of produced charm quark-pairs becomes negligible in the
final state, because charm-pair production is balanced by subsequent  absorption by the field.  
This dependence on the pulse width is also demonstrated on the right panel of Fig.~\ref{nt}. 
This figure clearly displays that charm quark-pair production is substantially enhanced in the cases of short pulse 
widths, and this enhancement has a maximum at $\tau \sim 0.1\sqrt{E_0}$.
In opposite to the heavy charm quark, light and strange quark-pair productions are increasing with the pulse width,
without any local maximum.

We have investigated the suppression factor and its dependence on pulse widths and quark masses.
Fig. \ref{gamma} summarizes our results. On the left panel 
the dependence on the pulse width is displayed. The strange to light ratio has a weak
dependence on the pulse width, its value is approaching  slowly  the 
asymptotic value of Schwinger limit (0.84) from below, similarly to U(1) gauge field~\cite{Prozorkevich:2004yp}. 
For charm quark this Schwinger limit is negligibly small, which value is reproduced by
our numerical calculation for very long pulse width.
On the other hand, at short pulse widths,
the relative charm production is surprisingly large, which does not follow any earlier
expectation. 
Considering charm to light and charm to strange ratios, only a slight difference
can be seen between them.
On the right panel we display the quark mass dependence of the suppression factor
for different pulse widths. For short pulse width the suppression factor is 
decreasing almost linearly with increasing quark mass value. For large pulse widths
we can see a very fast ($\sim \exp\{-m^2/E_0\}$) drop, which is consistent with
the Schwinger formula.

\section{Conclusion}

We have calculated light, strange and charm quark-pair production
in time-dependent SU(2) non-Abelian field. Applying a pulse-like
time evolution and investigating the influence of pulse width,
we observed that light and strange quark-pairs are produced as
we expected, approaching the Schwinger limit.
Charm quark-pairs followed this behaviour for large pulse widths.
However, for short pulses we did not see the expected charm 
suppression, connected to the large charm quark mass. 
Indeed, the large value of inverse temporal  width of the pulse,
overwhelming the mass of the heavy quark, $1 / \tau \gg m_c$,
determines the quark-pair production.
This finding could indicate the formation
of collective gluon field via enhanced heavy quark-pair production 
at RHIC. The issue of quantitative calculation of particle suppression 
factors for different quarks and comparison with the existing models 
will be addressed elsewhere.

\ack
This work was supported in part by  Hungarian OTKA Grants NK062044 and
NK077816, MTA-JINR Grant, and RFBR grant No. 08-02-01003-a.

\end{document}